\documentstyle[12pt,epsfig]{article} 
\newlength{\dinwidth}                       
\newlength{\dinmargin}                      
\newcommand{\ccaption}[2]{
  \begin{center}
   \parbox{0.85\textwidth}{
      \caption[#1]{\small{\it{#2}}}
                          }
  \end{center}
                         }
\def\beq{\begin{equation}}
\def\eeq{\end{equation}}
\def\beqn{\begin{eqnarray}}
\def\eeqn{\end{eqnarray}}
\def\journal#1#2#3#4{{\it #1 } {\bf #2} (19#3)#4}
\def\pl#1#2#3{{\it Phys. Lett. }{\bf #1}(19#2)#3}
\def\zp#1#2#3{{\it Z. Phys. }{\bf #1}(19#2)#3}

\def\prep#1#2#3{{\it Phys. Rep. }{\bf #1}(19#2)#3}
\def\pr#1#2#3{{\it Phys. Rev. }{\bf #1}(19#2)#3}
\def\np#1#2#3{{\it Nucl. Phys. }{\bf #1}(19#2)#3}

\def\hepph#1{{\tt hep-ph/#1}}

\def\sss{\scriptscriptstyle}
\def\as{\alpha_{\sss S}}         
\def\qq{{\sss Q\overline{Q}}}

\def\muf{\mu_{\sss F}}
\def\mur{\mu_{\sss R}}
\def\muo{\mu_0}

\def\aem{\alpha_{\rm em}}
\begin{document}

{\flushright{
        \begin{minipage}{4cm}
        ETH-TH/96-33 \hfill \\
        hep-ph/yymmxxx\hfill \\
        \end{minipage}        }

}

\begin{center}  \begin{Large} \begin{bf}
Prospects for heavy flavour photoproduction at 
HERA\footnote{~To appear in the proceedings of the workshop 
{\it Future Physics at HERA}, eds. G.~Ingelman, A.~De Roeck 
and R.~Klanner, DESY, Hamburg, 1996.}\\
  \end{bf}  \end{Large}
  \vspace*{5mm}
  \begin{large}
Stefano Frixione\\
  \end{large}
\vspace*{0.3cm}
Theoretical Physics, ETH, Z\"urich, Switzerland\\
\end{center}
\begin{quotation}
\noindent
{\bf Abstract:} I discuss few selected topics in heavy 
flavour photoproduction at HERA which require large integrated
luminosity in order to be experimentally investigated.
I present phenomenological predictions for bottom production.
As a possible application of measurements involving double-tagged 
charm events, I outline a method for the direct measurement of the gluon
density in the proton. The possibility of using charm data in
polarized electron-proton collisions to constrain the polarized
gluon density in the proton is also discussed.
\end{quotation}

Charm quarks are copiously produced at HERA. Total cross sections
in photoproduction have been measured~\cite{Derrick95,Aid96}, and
appear to be in reasonable agreement with next-to-leading order QCD 
calculations~\cite{Ellis89,Smith92,Frixione94a,NLOhadr}. 
Recently, the first results on single-inclusive 
distributions have been presented~\cite{Aid96}.
Although in substantial agreement with experiments, the theory displays
the tendency to undershoot the data. On the other hand, preliminary 
results by the ZEUS collaboration~\cite{ZEUS96} show sizeable
discrepancies in the comparison with next-to-leading order QCD,
especially in the pseudorapidity distribution. 

The limited statistics of the data prevents from any definite
conclusion on the capability of fixed-order QCD calculations to 
correctly describe charm photoproduction at the large center-of-mass
energies available at HERA. It has to be pointed out that the
resummation of logarithms which in certain regions of the phase space
may grow large and spoil the convergence of the perturbative
expansion, can not improve the comparison between theory and
experiments (see ref.~\cite{Eichler96} and references therein).
The luminosity upgrade of the HERA collider will allow to increase
the statistics of present measurements, and to 
perform new ones. The underlying theoretical picture will therefore
face a severe test. Detailed phenomenological predictions for 
total rates and single-inclusive distributions of charm quarks
in photoproduction at HERA have been available for some 
time~\cite{Riemersma92,Frixione95c,Frixione95b}. In the following,
I will deal with quantities whose measurement has
not yet been performed.

\section{Bottom production}

Due to the higher value of the quark mass, perturbative QCD predictions
for bottom production are more reliable than those for charm. 
In monochromatic photon-proton collisions, the pointlike component
has an uncertainty of a factor of 2 if all the parameters are varied
{\it together} in the direction that makes the cross section larger
or smaller. At $\sqrt{S_{\gamma p}}=100$~GeV, the lower and upper limits of
the pointlike component are 16 nb and 35 nb respectively, while at 
$\sqrt{S_{\gamma p}}=280$~GeV we get 41 nb and 101 nb~\cite{Frixione95c}. 
The hadronic component has larger uncertainties, but much smaller than 
for charm, since in bottom production the small-$x$
region is probed to a lesser extent than in charm production,
and the sensitivity of the result to the photon densities
is therefore milder; we get an uncertainty of a factor of 3
(to be compared with a factor of 10 in the case of charm).
The hadronic component can still be the dominant contribution
to the photoproduction cross section, if the gluon in the photon
is as soft as the LAC1 parameterization suggests.

\begin{figure}[thbp]
\centerline{\epsfig{figure=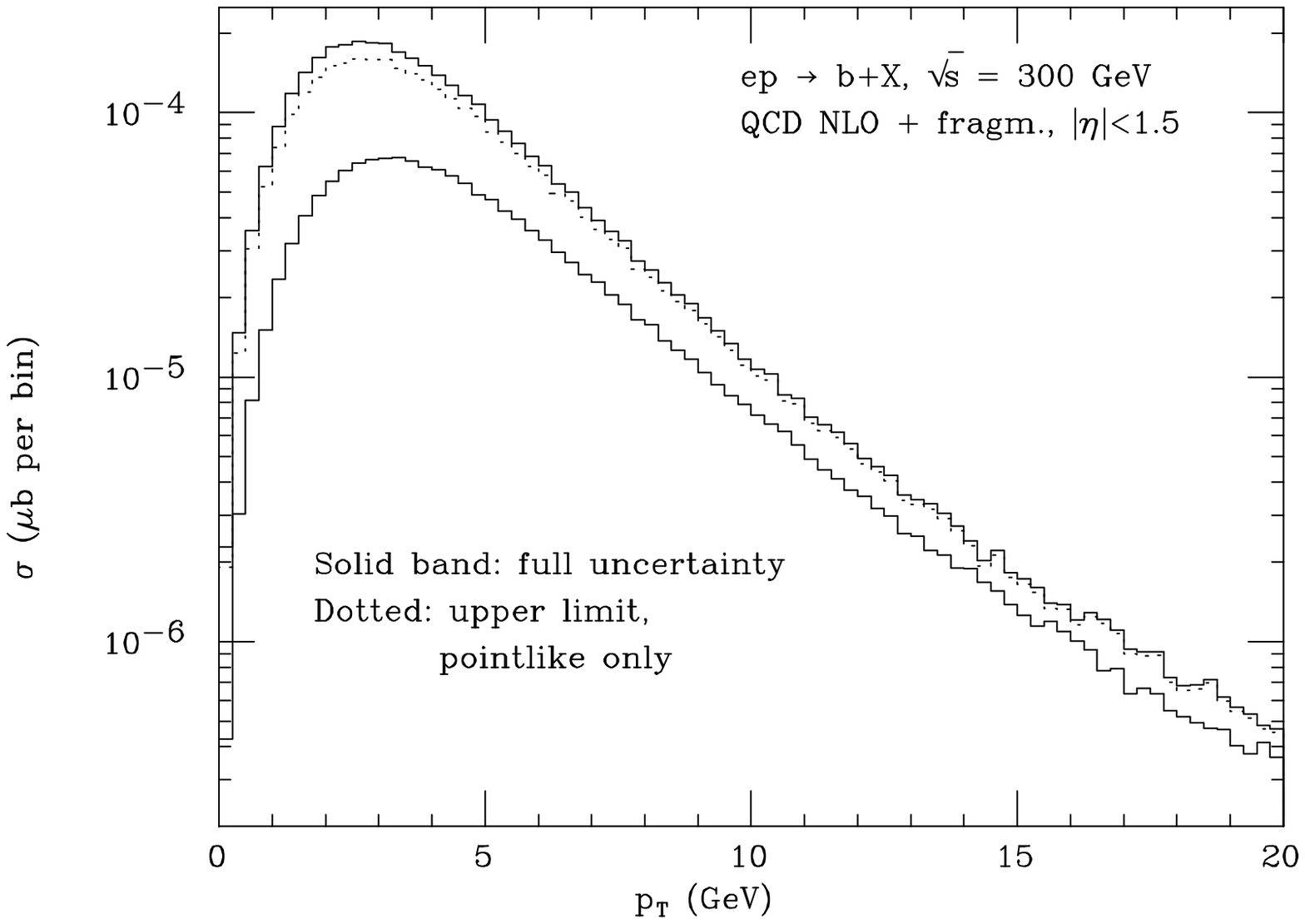,width=0.7\textwidth,clip=}}
\ccaption{}{ \label{f:b_pt_at_HERA}
Full uncertainty on the transverse momentum distribution for bottom 
electroproduction (Weizs\"acker-Williams approximation) with Peterson 
fragmentation and a pseudorapidity cut.}
\end{figure}
The bottom rates are about a factor of 200 smaller than the
charm ones. To perform a statistically significant study of bottom
production, the luminosity upgrade at HERA is necessary.
In any case, it is very likely that a comparison with the theory
could only be done by considering the electroproduction
(in the Weizs\"acker-Williams approximation) process.
In this case the sensitivity of the theoretical predictions to the 
input parameters is sizeably reduced, and a reliable 
comparison between theory and data can be performed.
For example, in electroproduction the hadronic component contribution 
to the total cross section is at most 75\% of the pointlike contribution, 
even if the LAC1 set is used. The most interesting results 
are however obtained when considering more exclusive quantities,
like single-inclusive distributions~\cite{Smith92,Frixione95b}. 
In particular, as shown in fig.~\ref{f:b_pt_at_HERA}, the 
transverse momentum of the bottom quark at HERA can be predicted 
by perturbative QCD quite accurately. It is clear that
even with the LAC1 set the hadronic component affects the prediction
only marginally; this fact is basically a consequence of the 
applied pseudorapidity cut. Figure~\ref{f:b_pt_at_HERA} can 
therefore be regarded as a reliable prediction of QCD for 
the $p_{\sss T}$ spectrum of $B$ mesons at HERA.
The comparison of this prediction with the data would be extremely
useful in light of the status of the comparison between theory 
and data for $b$ production at the Tevatron.

\section{The gluon density in the proton}

As discussed in ref.~\cite{Eichler96}, the experimental efficiency
for double-tagged charm events is quite low, and in order to study
fully-exclusive quantities, like the correlations between the charm
and the anticharm, a large integrated luminosity is mandatory.
The comparison between theoretical predictions and experimental
results for correlations constitutes the most stringent test for
the underlying theory, and it is therefore extremely interesting
for a complete understanding of the production mechanism of
heavy flavours. However, at present it is not particularly useful to
present detailed predictions for double-differential distributions,
since data will not be available for a long time. In the following,
I will therefore concentrate on a possible application of measurements
involving double-tagged charm events, namely the determination
of the gluon density in the proton ($f_g^{(p)}$).

At present, no {\it direct} measurement of $f_g^{(p)}$ has been 
performed. In principle, this quantity could be determined by 
investigating the exclusive properties of hard scattering
processes initiated by gluons. In practice, this procedure
is quite difficult; the data on direct photon production
and on inclusive jet production, which depends upon $f_g^{(p)}$
already at the leading order in QCD, are not as statistically
significant as DIS data are (DIS data allow a direct and accurate
determination of the quark densities). Direct photon
and inclusive jet data are used to constrain, in complementary
regions of $x$ and $Q^2$, the gluon density. Furthermore, in a 
next-to-leading order QCD evolution, $f_g^{(p)}$ affects the quark
densities through the Altarelli-Parisi equations, and therefore
has an impact on the description of DIS data on $F_2(x,Q^2)$
(for a detailed presentation of the determination of parton densities 
from a global QCD analysis, see for example refs.~\cite{Martin94,Lai96}).

A direct measurement of the gluon density is therefore highly
desirable. In ref.~\cite{Frixione93a} it was argued that 
charm production in high energy $ep$ collisions may help to
solve this problem. To proceed explicitly, I begin by writing the 
heavy-quark cross section at the leading order in the following form
\beq
\frac{d \sigma^{(0)}}{d y_{\qq} \, dM_{\qq}^2 }=
x_g\frac{d \sigma^{(0)}}{d x_g \, dM_{\qq}^2 }= \frac{1}{E^2}\;
f^{(e)}_{\gamma}(x_\gamma,\mu_0^2)
f^{(p)}_{g}(x_g,\muf^2)
\hat{\sigma}^{(0)}_{\gamma g}(M_{\qq}^2), 
\label{SigmaBorn}
\eeq
where $M_{\qq}$ is the invariant mass of the heavy-quark pair, and
$y_{\qq}$ is the rapidity of the pair in the electron-proton 
center-of-mass frame (I choose positive rapidities in the direction 
of the incoming proton). $E=\sqrt{S}$ is the electron-proton 
center-of-mass energy, and 
\beqn
x_\gamma&=&\frac{M_{\qq} }{E} \exp(-y_{\qq}),
\\
x_g&=&\frac{M_{\qq} }{E} \exp(y_{\qq}). 
\label{xgdef}
\eeqn
The function $f^{(e)}_{\gamma}$ is the Weizs\"acker-Williams 
function~\cite{WWfunction} (for a discussion on its use in production
processes involving heavy particles, see ref.~\cite{WW_heavy});
the explicit expression for the leading-order
cross section $\hat{\sigma}^{(0)}_{\gamma g}$ can be found in
ref.~\cite{Ellis89}. The factorization and renormalization scales
($\muf$ and $\mur$) are set equal to $2\muo$ and $\muo$ respectively,
where $\muo$ is a reference scale; when studying correlations, it
is customary to choose~\cite{Frixione94a}
\beq
\muo=\sqrt{(p_{\sss T}^2+\bar{p}_{\sss T}^2)/2+m^2},
\eeq
where $p_{\sss T}$ and $\bar{p}_{\sss T}$ are the transverse momenta 
of the heavy quark and of the heavy antiquark respectively.

Assuming that the left-hand side of eq.~(\ref{SigmaBorn}) is 
identified with the data, the equation can be inverted, and we can 
get a first determination of $f^{(p)}_g$:
\beq
f^{(0)}_g(x_g,\muf^2)=
x_g\frac{d \sigma^{\rm data}}{d x_g \, dM_{Q\bar {Q}}^2 } \frac{E^2}
{f^{(e)}_{\gamma}(x_\gamma,\mu_0^2)
\hat{\sigma}^{(0)}_{\gamma g}(M_{\qq}^2) }.
\eeq
The inclusion of radiative corrections does not pose any problem.
I write the full cross section as 
\beq 
x_g \frac{d \sigma}{d x_g \, dM_{\qq}^2 }=
\frac{1}{E^2}\;f^{(e)}_{\gamma}(x_\gamma,\mu_0^2)
f^{(p)}_{g}(x_g,\muf^2)
\hat{\sigma}^{(0)}_{\gamma g}(M_{\qq}^2)
+\Delta(f^{(p)}_g,x_g,M_{\qq}^2),
\label{SigmaNL}
\eeq
where $\Delta$ represents all the radiative effects. In $\Delta$ I have also
indicated explicitly the functional dependence upon the gluon density in the
proton. The light quarks, which enter at the next-to-leading order via the
$\gamma q\to q Q\bar{Q}$ process, give a small contribution (less than 5\% for
all values of $x_g$ and $M_\qq$ considered here). 
I now write $f^{(p)}_g$ as
\beq
f^{(p)}_g(x,\muf^2)=f^{(0)}_g(x,\muf^2)+f^{(1)}_g(x,\muf^2),
\eeq
where the second term is the next-to-leading order correction, and plug 
it back into eq.~(\ref{SigmaNL}). Then
\beq
f_g^{(1)}(x_g,\muf^2)=-\frac{E^2\Delta(f^{(0)}_g,x_g,M_{\qq}^2)}
{f^{(e)}_\gamma(x_\gamma,\mu_0^2)
\;\hat{\sigma}^{(0)}_{\gamma g}(M_{\qq}^2) }.
\label{f1correction}
\eeq
I have neglected the $f_g^{(1)}$ piece contained in the $\Delta$ term,
the corresponding contribution being of order $\aem\as^3$.
It could also be easily incorporated by iterating
eq.~(\ref{f1correction}), using the full gluon density in the
right-hand side.
\begin{figure}
\centerline{\epsfig{figure=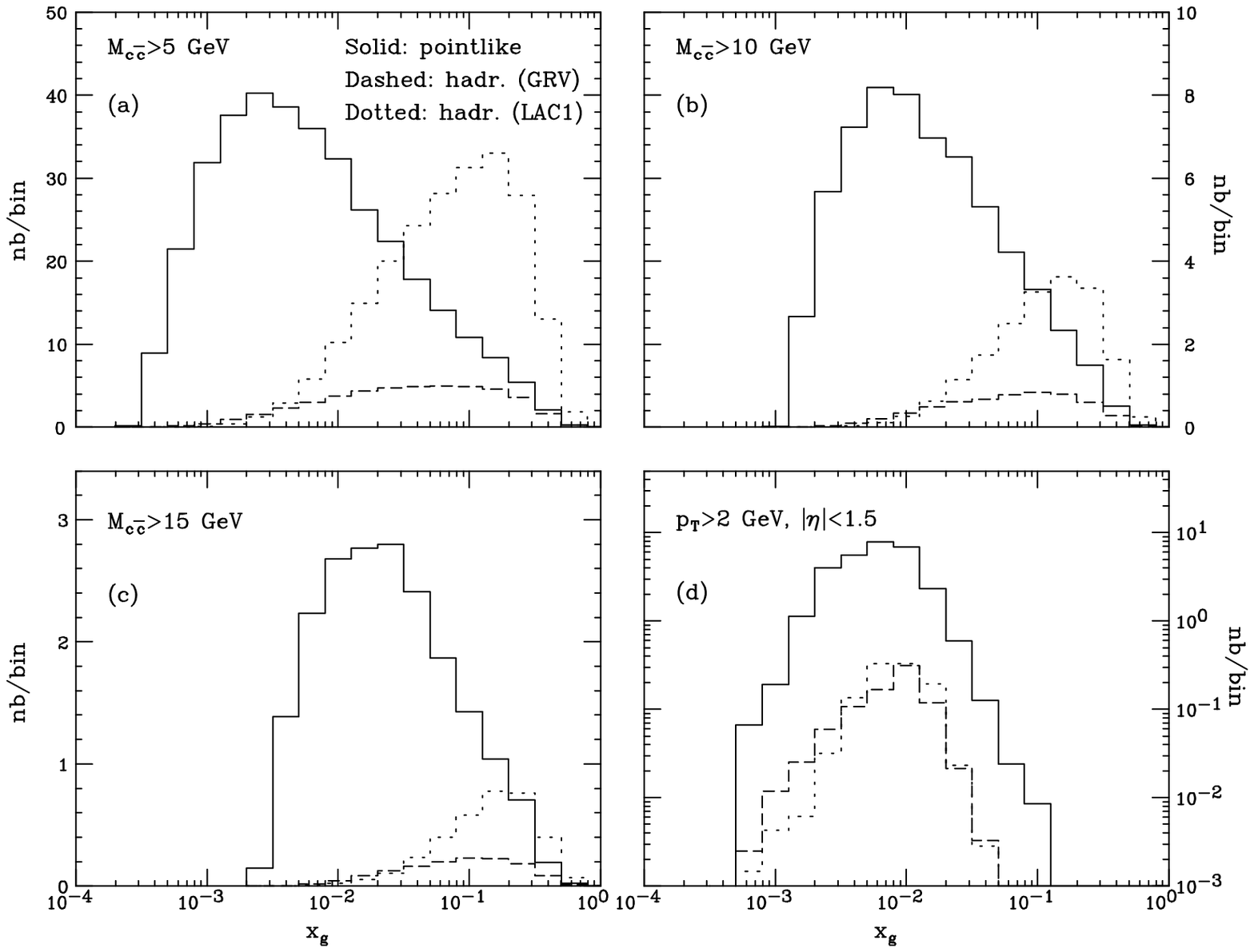,width=0.75\textwidth,clip=}}
\ccaption{}{ \label{f:xgdistr}
$x_g$ distribution in $ep$ collisions (Weizs\"acker-Williams
approximation) at HERA, for $m_c=1.5$~GeV. The 
proton parton density set adopted is MRSG.}
\end{figure}

The charm cross section receives a large
contribution from the hadronic component, which was neglected in
the above derivation. In order to extract the
gluon density in the proton with the method previously outlined,
we have to consider only those kinematical regions where the
hadronic component is suppressed by the dynamics. 
I study this possibility in figure~\ref{f:xgdistr},
where I present the next-to-leading order QCD
predictions for the variable $x_g$, defined in eq.~(\ref{xgdef}),
in electron-proton collisions at $\sqrt{S}=314$~GeV. The partonic
densities in the proton are given by the MRSG set, while both the LAC1 
and GRV-HO sets for the photon are considered, in order
to account for the uncertainty affecting the gluon density in the photon. 
Figs.~\ref{f:xgdistr}(a)-\ref{f:xgdistr}(c) show the effect of 
applying a cut on the invariant mass of the pair. Even in the
case of the smallest invariant-mass cut, there is a region of small 
$x_g$ where the hadronic component is negligible with respect to the 
pointlike one. When the invariant-mass cut is increased, the hadronic 
component can be seen to decrease faster than the pointlike one.
This is due to the fact that, for large invariant masses of the pair,
the production process of the hadronic component is suppressed by the small 
value of the gluon density in the photon at large $x$. By pushing
the invariant-mass cut to 20 GeV, it turns out that the pointlike
component is dominant over the hadronic one for $x_g$ values
as large as $10^{-1}$. The conclusion can be drawn that the
theoretical uncertainties affecting the charm cross section,
in the range $10^{-3}<x_g<10^{-1}$, are small
enough to allow for a determination of the gluon density in the proton
by using invariant-mass cuts to suppress the hadronic component.
In a more realistic configuration, like the present one of 
the detectors at HERA, additional cuts are applied to the data.
Fig.~\ref{f:xgdistr}(d) shows the effect on the $x_g$ distribution
due to a small-$p_{\sss T}$ and a pseudorapidity cut, applied 
to both the charm and the anticharm. In this case, 
even without an invariant-mass cut, the pointlike
component is dominant in the whole kinematically accessible range.
Taking into account experimental efficiencies~\cite{Eichler96},
a statistically significant measurement of the gluon density 
requires an integrated luminosity of at least 250 pb$^{-1}$.

\section{Polarized $ep$ collisions}

It is conceivable that in the future the HERA collider will be operated 
in a polarized mode. The heavy flavour cross section in polarized $ep$
scattering is dependent already at the leading order in QCD
upon the polarized gluon density in the proton, $\Delta g^{(p)}$.
Therefore, data on charm production could be
used to directly measure $\Delta g^{(p)}$, as previously
shown for the unpolarized case. In practice, the situation
for the polarized scattering is much more complicated.
First of all, a full next-to-leading order calculation is
not available for the partonic processes relevant for
polarized heavy flavour production. Furthermore, there is no experimental
information on the partonic densities in the polarized photon.
It is reasonable, however, to think that charm production
at the HERA collider in the polarized mode can help in constraining
the polarized gluon density in the proton. This possibility
was first suggested in ref.~\cite{Deltag}, and recently reconsidered 
in refs.~\cite{Frixione96b,Stratmann96}.

The next-to-leading and higher order corrections to the polarized
cross section are expected to be sizeable, therefore casting 
doubts on the phenomenological relevance of leading-order predictions.
To overcome this problem, one possibility is to present 
predictions~\cite{Frixione96b} for the ratio $\Delta\sigma/\sigma$
(asymmetry), where $\sigma$ is the unpolarized cross section and
\beq
\Delta\sigma=\frac{1}{2}
\left(\sigma^{\uparrow\uparrow}-\sigma^{\uparrow\downarrow}\right).
\eeq
Here $\sigma^{\uparrow\uparrow}$ and $\sigma^{\uparrow\downarrow}$ 
are the cross sections for $c\bar{c}$ production with parallel and
antiparallel polarizations of the incoming particles respectively.
One might expect that the effect of the radiative corrections 
approximately cancels in the ratio. It has to be stressed that, 
for consistency reasons, the unpolarized cross section $\sigma$ 
appearing in the asymmetry must be calculated at the leading order, 
as the polarized one.
\begin{figure}[ptbh]
  \begin{center}
    \mbox{
      \epsfig{file=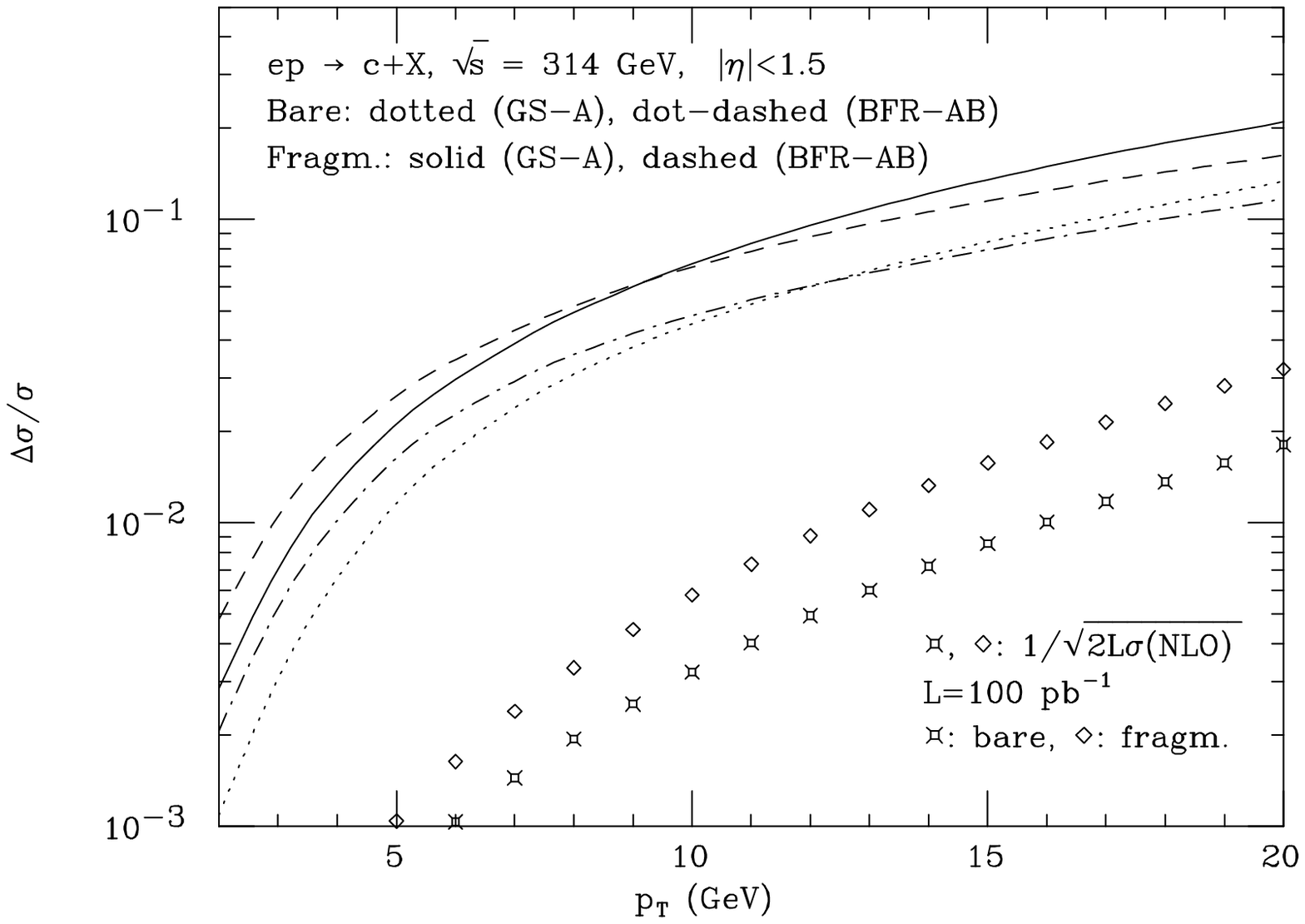,width=0.70\textwidth}
      }
  \ccaption{}{\label{f:ptel}
Asymmetry cross section versus transverse momentum in polarized $ep$ 
collisions (Weizs\"acker-Williams approximation) at $\sqrt{S}=314$~GeV. 
The minimum observable asymmetry, computed at next-to-leading order, 
is also displayed.}
  \end{center}
\end{figure}

The next-to-leading order value of $\sigma$ can then be
used to estimate the sensitivity of the experiment. A rough estimate of the 
minimum value of the asymmetry observable at HERA can be obtained by requiring 
the difference between the numbers of events with parallel and antiparallel
polarizations of the initial state particles to be larger than the 
statistical error on the total number of observed events. This gives
\beq
\left[\frac{\Delta\sigma}{\sigma}\right]_{min}\simeq
\frac{1}{\sqrt{2\sigma{\cal L}\epsilon}},
\label{minassmtr}
\eeq
where $\cal L$ is the integrated luminosity and the factor $\epsilon$
accounts for the experimental efficiency for charm identification
and for the fact that the initial beams are not completely polarized.
This procedure can be applied to total cross sections, as well as
to differential distributions; in this case, the values of
$\sigma$ and $\Delta\sigma$ have to be interpreted as {\it cross
sections per bin} in the relevant kinematical variable.

In ref.~\cite{Frixione96b} it was shown that total cross section 
asymmetries for the pointlike component are quite small in absolute 
value, and can be measured only if $\epsilon$ is equal or 
larger than 1\% (0.1\%), assuming ${\cal L}=$~100~pb$^{-1}$
(1000~pb$^{-1}$). Therefore, even with a vertex detector (see
ref.~\cite{Eichler96}), it appears to be unlikely that this
kind of measurements will be performed at HERA. Furthermore,
in ref.~\cite{Stratmann96} a rough estimate of the hadronic contribution
to the polarized cross section has been given, assuming polarized parton 
densities in the photon to be identical to zero or to the unpolarized 
densities to get a lower and an upper bound on the cross section.
It was found that a non-negligible contamination of the pointlike
result might indeed come from the hadronic process.
The situation clearly improves when considering more exclusive
quantities; in ref.~\cite{Frixione96b} it was found that at moderate
$p_{\sss T}$ values the asymmetry for the pointlike component
can be rather large, well above the minimum observable 
value (in this region, the experimental efficiency is
sizeable~\cite{Eichler96}); this is shown in fig.~\ref{f:ptel}. In 
ref.~\cite{Stratmann96} it was argued that the hadronic component
should have a negligible impact in this case.
I conclude that, with an integrated luminosity of 100~pb$^{-1}$, 
charm data in high-energy polarized $ep$ collisions
will help in the determination of the polarized gluon density
in the proton. In order to distinguish among different parameterizations
for $\Delta g^{(p)}$, a larger luminosity is likely to be needed.

\section{Conclusions}

I have discussed few selected topics in heavy flavour physics
which will become of practical interest at HERA after that the
planned upgrades of the machine will be carried out. With an
integrated luminosity ${\cal L}=$~100~pb$^{-1}$, about $10^5$
bottom quarks are predicted by QCD to be produced in $ep$
collisions at $\sqrt{S}=300$~GeV. If the experimental efficiency
for $B$-meson identification will be large enough, this will
provide with the possibility of a detailed study of the
bottom production mechanism, and of an interesting comparison
with the results at the Tevatron. In order to study charm-anticharm
correlations, ${\cal L}$ must be equal to or larger than 250~pb$^{-1}$.
As a possible application of measurements involving double-tagged 
charm events, I presented a method for the direct measurement of the
gluon density in the proton. Charm data in polarized $ep$
collisions could also be used to constrain the {\it polarized}
gluon density in the proton. In this case, an integrated
luminosity of at least 100~pb$^{-1}$ is required.

\vspace*{0.5cm}
\noindent {\bf Acknowledgements:} The financial support by the 
Swiss National Foundation is acknowledged.

\end{document}